\documentclass[aps,twocolumn,nofootinbib,showkeys,showpacs,superscriptaddress]{revtex4-1}

\usepackage{amssymb}
\usepackage{amsmath}
\usepackage{amstext}
\usepackage{amsfonts}
\usepackage{bbold}
\usepackage{bm}
\usepackage{graphics}
\usepackage{mathtools}
\usepackage{lmodern}
\usepackage{graphicx}
\usepackage{hyperref}
\usepackage{xcolor}
\usepackage[export]{adjustbox}
\usepackage{soul}
\usepackage{microtype}
\usepackage{enumerate}

\newcommand{\sss}{\scriptscriptstyle}

\begin{document}

\title{Effective field theory for acoustic and pseudo-acoustic phonons in solids} 

\author{Angelo~Esposito}
\affiliation{Theoretical Particle Physics Laboratory (LPTP), Institute of Physics, EPFL, 1015 Lausanne, Switzerland}

\author{Emma~Geoffray}
\affiliation{Institute for Theoretical Physics, Heidelberg University, D-69120 Heidelberg, Germany}

\author{Tom~Melia}
\affiliation{Kavli Institute for the Physics and Mathematics of the Universe (WPI), UTIAS, The University of Tokyo, Kashiwa, Chiba 277-8583, Japan}

\begin{abstract}
We present a relativistic effective field theory for the interaction between acoustic and gapped phonons in the limit of a small gap. We show that, while the former are the Goldstone modes associated with the spontaneous breaking of spacetime symmetries, the latter are pseudo-Goldstones associated with some (small) explicit breaking. We hence dub them ``pseudo-acoustic'' phonons. In this first investigation, we build our effective theory for the cases of one and two spatial dimensions, two atomic species, and assuming large distance isotropy. As an illustrative example, we show how the theory can be applied to compute the total lifetime of both acoustic and pseudo-acoustic phonons. This construction can find applications that range from the physics of bilayer graphene to sub-GeV dark matter detectors. 
\end{abstract}


\maketitle


\section{Introduction}

Many properties of solids are dictated by the dynamics of their simplest collective excitations: the phonons. These are localized vibrational modes that, when characterized by wavelengths much larger than the atomic spacing, can be described in terms of quasiparticles. In a solid with a single atom per unit cell the phonons' dispersion relation is gapless; i.e., when its wave vector vanishes so does its frequency. In this case one talks about ``acoustic'' phonons. However, for more complicated (and common) solids, some phonons can be gapped, with a frequency that tends to a finite positive value at zero wave vector. When the gap is comparable to or larger than the maximum frequency of the acoustic phonons, these modes are typically called ``optical'' phonons.

It is well known that acoustic phonons are Goldstone bosons associated with the spontaneous breaking of spatial translations induced by a solid background~\cite{Leutwyler:1996er}.
Taking this idea as a starting point, we see that recent years have witnessed the development of relativistic effective field theories (EFTs), based on symmetry breaking and its consequences, applied to the study of collective excitations in different media (see~\cite{Son:2002zn,Endlich:2012pz,Nicolis:2015sra} and references therein). An EFT description of the phonons, organized in a low-energy/long wavelength expansion,  has the advantage of being universal; i.e., it does not rely on the often complicated microscopic physics, up to a finite number of effective coefficients. The latter must be obtained from experiment or determined in other ways, as, for example, density functional theory (DFT) calculations---see, e.g.,~\cite{dft1,PhysRevB.78.165421,dft3,dft2,dft4}. 
Such an EFT approach has already proven to be useful to a number of phenomenologically relevant problems, covering a wide range of fields, from the physics of $^4$He to cosmology (see, e.g.,~\cite{Golkar:2014wwa,Horn:2015zna,Crossley:2015evo,Nicolis:2017eqo,Rothstein:2017niq,Moroz:2018noc,Alberte:2018doe,Esposito:2018sdc,Acanfora:2019con,Caputo:2019xum,Caputo:2019cyg}).

\begin{figure}[t]
\centering
\includegraphics[width=\columnwidth]{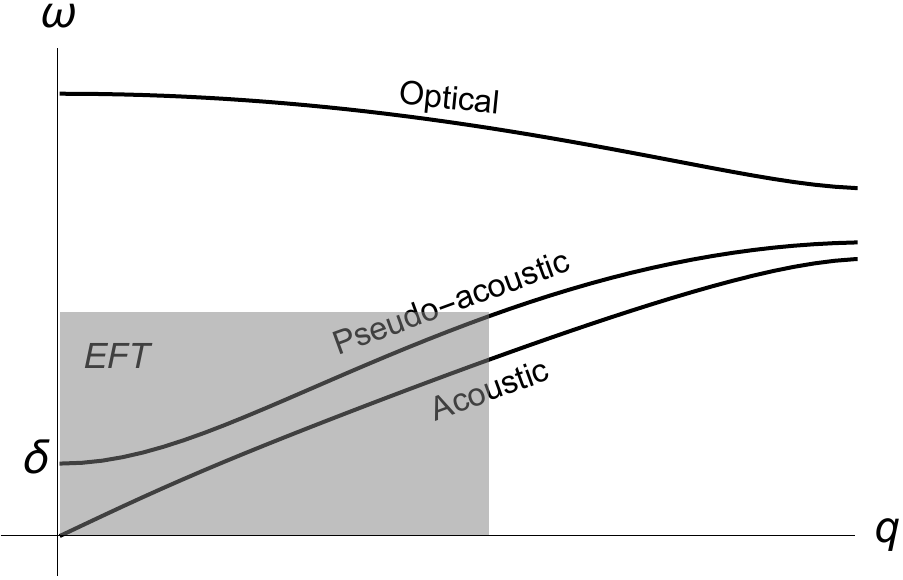}
\caption{Schematic representation of the energy of different phonon modes as a function of momentum. The  shaded region indicates where our EFT is valid.} \label{fig:spectrum}
\end{figure}

In this paper we develop a new relativistic EFT for the description of the interactions of acoustic and gapped phonons in a solid, in the regime where the gap is small compared to the typical frequency characterizing the microscopic system. For reasons that will be clear soon, we dub these collective excitations ``pseudo-acoustic'' phonons. See Fig.~\ref{fig:spectrum} for a schematic representation of the phonon spectrum.
To the best of our knowledge, no bottom-up effective description of pseudo-acoustic phonons has been presented thus far.\footnote{See~\cite{opticalEFT1,opticalEFT2} for a proposal on how to describe optical phonons in a quantum field theory language. 
For the inclusion of explicit breaking of translations in hydrodynamic transport as well as nonrenormalizable field theories see, e.g.,~\cite{Delacretaz:2016ivq,Delacretaz:2017zxd,Delacretaz:2019wzh,Amoretti:2018tzw,Amoretti:2019cef,Amoretti:2019kuf,Musso:2018wbv,Musso:2019kii}.}

This construction can be applied to any number of spatial dimensions and any number of atomic species within the solid. However, as we will show, we expect a small gap for the pseudo-acoustic phonons to arise when the different species are weakly coupled to each other. {Such an instance is realized, for example, in few-layer materials such as graphene~\cite{PhysRevB.78.165421,PhysRevB.78.113407,xxx,doi:10.1063/1.4735246,PhysRevB.83.235428,Nika_2012,PhysRevB.88.035428,nika2017phonons}, hexagonal boron nitride~\cite{michel2011phonon,d2017length}, or a combination of the two~\cite{gholivand2017phonon}. Indeed, we expect pseudo-acoustic phonons to arise in approximately two-dimensional materials every time the layers are weakly coupled to each other. In three dimensions, this situation might be harder to envision. Nevertheless, we notice that the spectrum of both graphite and three-dimensional hexagonal boron nitride exhibits phonon branches that closely resemble pseudo-acoustic modes~\cite{kern1999ab,wirtz2004phonon,reich2005resonant,michel2008theory,gholivand2017phonon}.}

In this first study, we focus on solids that are both homogeneous and isotropic at large distances. While the first property is always true, the second one is a simplifying assumption. The extension of our EFT to the case of solids which preserve only discrete rotations at large distances is straightforward, but tedious. 

In constructing the EFT for acoustic and pseudo-acoustic phonons we impose relativistic Lorentz invariance.
Although this might not be common in solid-state physics, there are reasons why this approach is worth exploring. First, it is technically easier to impose the Lorentz symmetry than the nonrelativistic Galilei one, by simply contracting covariant indices. Moreover, given that the Lorentz group is more fundamental, one is always free to require invariance under it, and hence allow the EFT to include relativistic effects on the phonon dynamics. The nonrelativistic limit can always be taken afterward, by reintroducing the speed of light with simple dimensional analysis and formally sending it to infinity (see, e.g.,~\cite{Esposito:2018sdc}). That being said, if the system of interest is nonrelativistic, there is no conceptual obstruction in imposing invariance under Galilei boosts right away.
Finally, the formulation of an EFT in relativistic language  also provides a useful connection to high energy physics, for example, enabling observables to be computed using techniques more familiar to this community. 

{One promising application of this is in the use of EFTs for describing the interactions of sub-GeV dark matter with a detector. A number of recent proposals utilize dark matter scattering off phonons in various materials as a detection mechanism; see, e.g.,~\cite{Hochberg:2016ajh,Guo:2013dt,Schutz:2016tid,Knapen:2016cue,Acanfora:2019con,Caputo:2019cyg,Caputo:2019xum,Baym:2020uos,Knapen:2017ekk,Campbell-Deem:2019hdx,Griffin:2018bjn,Griffin:2019mvc,Trickle:2019nya,Arvanitaki:2017nhi,Bunting:2017net,Chen:2020jia,Cox:2019cod}. For low-mass dark matter with a large de Broglie wavelength, an effective description of the coupling between dark matter and collective modes in  the detector is necessary, since the microscopic features are not resolved. The EFT approach we follow can be extended to capture the most general dark matter--detector interaction, which is necessary for studying generic rates and differential rates of dark matter scattering. Furthermore, curious cancellation mechanisms  have been observed in low-mass dark matter scattering rates in liquid helium that are transparent in the EFT~\cite{Knapen:2016cue,Caputo:2019cyg}. Similar cancellations have been observed in traditional calculations of dark matter scattering off gapped and acoustic phonons in crystals~\cite{Cox:2019cod,Campbell-Deem:2019hdx}.  The EFT we introduce could provide new insight into these mechanisms. It could also be utilized to calculate rates of dark matter scattering off previously unconsidered pseudo-acoustic phonons,  which lie in an interesting region between acoustic and optical phonons (the former having a larger scattering rate, the latter being gapped and thus having more favorable kinematics to match those of dark matter~\cite{Knapen:2017ekk}); this could in turn provide the impetus to consider materials that have pseudo-acoustic phonons as dark matter detectors.}

{The EFT of pseudo-acoustic phonons could also find application in ``solid'' theories of inflation;  see~\cite{Endlich:2012pz}.}

\section{EFT for Acoustic and Pseudo-Acoustic Phonons}

 We now describe the EFT for pseudo-acoustic and acoustic phonons, focusing on the simple cases of one and two spatial dimensions, as well as two atomic species.

\subsection*{1D case}

Consider for a moment a monoatomic solid~\cite{Endlich:2012pz,Nicolis:2015sra,Esposito:2018sdc}. Its volume elements can be labeled with a single scalar field, $\phi(x)$---the comoving coordinate---which at equilibrium can be taken to be proportional to the physical spatial coordinate, $\langle\phi(x)\rangle=\alpha x$, where $\alpha$ is a constant determining the degree of compression/dilation of the solid~\cite{Landau:1986aog}. From an EFT viewpoint this vacuum expectation value breaks boosts and spatial translations. Since all solids are homogeneous at large distances, one also postulates an internal $U(1)$ shift symmetry, $\phi\to\phi+c$, which is broken together with part of the Poincar\'e group
down to time translations and a diagonal $U(1)$, i.e., $ISO(1,1)\times U(1)\to \mathbb{R}_t\times U(1)$. It is this last unbroken $U(1)$ that one uses to define large distance homogeneity.

The fluctuation of the comoving coordinate around its equilibrium configuration, $\phi(x)=\alpha x+\pi(x)$, is the Goldstone boson associated with the broken symmetries, corresponding to the phonon of the solid. As the breaking is spontaneous, the phonon dynamics must be described via a Lagrangian that is invariant under the full initial group. In the long wavelength limit (i.e., at lowest order in a derivative expansion), the only quantity that is invariant under both Poincar\'e transformations and the internal shift symmetry is $X=\partial_\mu\phi\partial^\mu\phi$, and the most general Lagrangian is $F(X)$, with $F$ an a priori generic function. Upon inspecting the stress-energy tensor of the theory, one finds that $F$ is simply minus the energy density~\cite{Endlich:2012pz}. For a strongly coupled system its analytical expression is hard (or even impossible) to compute, and one must obtain it from experimental or numerical data.

Expanding the action in small fluctuations one obtains all possible interactions for the acoustic phonon which, being a Goldstone boson, is gapless---for details see, e.g.,~\cite{Endlich:2012pz}.

\vspace{1em}

Let us now consider a second atomic species in our solid. One can introduce two comoving coordinates, $\phi_{A,B}(x)$, one for each species, featuring two independent shift symmetries. At equilibrium both of them are proportional to space, $\langle\phi_A(x)\rangle=\alpha x$ and $\langle\phi_B(x)\rangle=\beta x$, and the symmetry breaking pattern is then $ISO(1,1)\times U_A(1)\times U_B(1)\to \mathbb{R}_t\times U(1)$. 
Despite the number of broken generators, the system features only two Goldstones, as dictated by the inverse Higgs constraints~\cite{Ivanov:1975zq,Watanabe:2013iia,Nicolis:2013lma}. 
In the following we set $\alpha=\beta=1$ for simplicity.\footnote{{At equilibrium the parameters $\alpha$ and $\beta$ are fixed and specified by either minimization of the solid's free energy or by boundary conditions (external pressure, compression, etc.). It is then always possible to set $\alpha=\beta=1$ by a redefinition of the comoving coordinates. Although  varying the solid background amounts to changing $\alpha$ and $\beta$, we can keep track of it via the dependence of the Lagrangian on the invariants in Eq.~\eqref{eq:invariants}.}}

Analogously to the monoatomic case, at lowest derivative order one can build three quantities that are invariant under Poincar\'e transformations and the two internal $U(1)$'s, i.e.,
\begin{align} \label{eq:invariants}
\begin{split}
    X_1=\partial_\mu\phi_A&\partial^\mu\phi_A\,, \qquad X_2=\partial_\mu\phi_B\partial^\mu\phi_B\,, \\
    & X_3=\partial_\mu\phi_A\partial^\mu\phi_B\,.
\end{split}
\end{align}
The most general action will depend on all these invariants, describing two interacting acoustic phonons, both gapless.

This is, however, not the end of the story. It is now possible to build one more quantity, $\Delta=(\phi_A-\phi_B)^2$, which explicitly breaks the initial $U_A(1)\times U_B(1)$ but preserves their final diagonal combination.\footnote{
We define $\Delta$ so that the theory is invariant under $\phi_{A,B}\to-\phi_{A,B}$ while being analytic around $\pi_{A,B}=0$.
} This operator generates a gap for one of the two degrees of freedom, which then becomes a \emph{pseudo}-Goldstone boson---hence the name pseudo-acoustic phonon.\footnote{Note that, despite the presence of a gap (albeit small), these phonons are usually still called ``acoustic'' in the literature; see, e.g.,~\cite{Nika_2012,PhysRevB.88.035428,nika2017phonons}. We stress, however, the conceptual difference between the two: while standard acoustic phonons are Goldstone bosons, pseudo-acoustic phonons are not.} If such a gap is smaller than the UV cutoff, the pseudo-acoustic phonon can still be treated perturbatively within the EFT, very much like pions in QCD~\cite{Donoghue:1992dd}.
This means that, while the $A$ and $B$ solids can be separately arbitrarily strongly coupled at the microscopic level, we expect this regime to be achieved when they are weakly coupled to each other. 
A priori, the most general Lagrangian that incorporates a small explicit breaking (and hence a small gap) is 
\begin{align}
F(X_i,\Delta)=f(X_i) + \delta f(X_i,\Delta)\,,
\end{align}
with $\delta f\ll f$. However, for the most common systems we expect that if the two solids are weakly coupled to each other, all interactions between the phonons of the two sectors will be weak, i.e.,
\begin{align} \label{eq:Fweak}
    F_{\text{weak}}(X_i,\Delta) = f_A(X_1) + f_B(X_2) + \delta f(X_i,\Delta)\,,
\end{align}
where $f_{A,B}$ are (minus) the energy densities of the two solids in the limit where they are exactly decoupled.

Expanding the action
$S=\int d^2x \, F\big(X_i,\Delta\big)$ in small fluctuations, up to cubic order one gets
\begin{align} \label{eq:S21D}
\begin{split}
    S&\supset\int d^2x\bigg[-g_{(\alpha\beta)}\dot\pi_\alpha\dot\pi_\beta+g^\prime_{(\alpha\beta)}\partial_x\pi_\alpha\partial_x\pi_\beta \\
    & \quad +g_{\sss \Delta}\big(\pi_A-\pi_B\big)^2 + y_{(\alpha\beta\gamma)}\,\partial_x\pi_\alpha\partial_x\pi_\beta\partial_x\pi_\gamma \\
    & \quad -y_{(\alpha\beta)\gamma}^\prime\,\dot\pi_\alpha\dot\pi_\beta\partial_x\pi_\gamma+y^{\prime\prime}_\alpha\big(\pi_A-\pi_B\big)^2\partial_x\pi_\alpha\bigg]\,,
\end{split}
\end{align}
where $\alpha,\beta,\gamma=A,B$ and $(\,\cdots)$ represents symmetric indices. The effective couplings, $g$ and $y$, in terms of the derivatives of $F$ can be found in Appendix~\ref{app:couplings}. In general, the coefficients of the quadratic terms depend on derivatives up to the second, those of the cubic ones up to the third, and so on. The spectrum of the theory is obtained by looking at the eigenmodes of the quadratic part of the action. For small momenta one gets the following dispersion relations for acoustic and pseudo-acoustic phonons:
\begin{align} \label{eq:dispersion1D}
    \omega_\text{ac}^2 &= c_s^2 q^2\,, \qquad \; \omega_\text{ps}^2 = \delta^2 +\gamma q^2\,,
\end{align}
where the gap is given by $\delta^2\equiv\frac{g_{\sss AA}+g_{\sss BB}+2g_{\sss AB}}{g_{\sss AA}g_{\sss BB}-g_{\sss AB}^2}g_{\sss \Delta}$.
The expressions for $c_s$ and $\gamma$ are also reported in Appendix~\ref{app:couplings}. The gap of the pseudo-acoustic phonon indeed goes to zero with vanishing $g_{\sss \Delta}$, with this being the only parameter encoding the dependence on explicit breaking at quadratic order.

\vspace{1em}

To make contact with physical systems, let us show how the most general theory~\eqref{eq:S21D} can describe two examples of a linear diatomic chain. We focus on the spectrum only. 

\begin{enumerate}[\hspace{1em}a.]

    \item
    
     \emph{Two noninteracting monoatomic chains.}---A system of this kind is described, as already mentioned, by a Lagrangian as in Eq.~\eqref{eq:Fweak} in the $\delta f\to0$ limit. 
     This implies that $g_{\sss AB}=g'_{\sss AB}=g_{\sss \Delta}=0$, and one obtains two gapless modes with two generically different sound speeds, $c_{\sss A}^2=g_{\sss AA}^\prime/g_{\sss AA}$ and $c_{\sss B}^2=g_{\sss BB}^\prime/g_{\sss BB}$. Note that since the $\bm{q}\to0$ and $\delta\to0$ limits do not commute, one cannot get the above sound speeds as a limit of the dispersion relations~\eqref{eq:dispersion1D}.
    
   \item
     \emph{Two identical chains with weak coupling between them.}---As the two separate chains are identical, the system is obtained from action~\eqref{eq:Fweak} imposing $f_A=f_B$ and symmetry under $X_1\leftrightarrow X_2$, which implies that $g_{\sss AA}^{(\prime)}=g_{\sss BB}^{(\prime)}$ and $g_{\sss AB}^{(\prime)}=0$. In this case the small gap survives but one obtains $\gamma\simeq c_s^2$. 
    
\end{enumerate}

Let us close this section by briefly discussing how the effective couplings of the theory can be determined from the static properties of the solid, e.g., by experiment or numerical simulations. It is clear that the structure of action~\eqref{eq:S21D} could also have been found by simply writing down all possible interactions compatible with the unbroken symmetries. Nevertheless, to express the couplings in terms of derivatives of the Lagrangian with respect to the invariants allows one to determine them in terms of the nonlinear stress-strain curve of the solid~\cite{Alberte:2018doe}. Imagine for example,  statically stretching or compressing only one of the solids, say solid $A$, while keeping the other at its equilibrium configuration. This corresponds to exciting a time-independent mode $\pi_A(x)$ while keeping $\pi_B=0$, i.e., a deformation of solid $A$. At linear order in the deformation $\pi_A$, this induces a variation in $X_1$, $X_3$, while $X_2$ remains unchanged. Exciting a mode $\pi_B(x)$ has the same effect but with $X_1\leftrightarrow X_2$. If instead we statically deform the two solids in opposite directions, $\pi_A(x)=-\pi_B(x)$, this will generate a variation in $X_1$, $X_2$ but not in $X_3$. 

Recalling that the Lagrangian, $F(X_i,\Delta)$, is minus the energy density of the solid, one deduces that the effective couplings can be obtained by studying the nonlinear response of the energy density under the mechanical deformations described above, as can be done, say, using DFT techniques~\cite{dft1,PhysRevB.78.165421,dft3,dft2,dft4}. For example, by measuring the linear change in the free energy following the deformations described above, one can determine the first derivatives of the Lagrangian with respect to the $X_i$ invariants. To obtain higher derivatives of the Lagrangian with respect to $X_i$, as well as the dependence on $\Delta$, one can study the nonlinear response.


\subsection*{2D case}

Building on the previous section, we now describe the case of a two-dimensional diatomic solid. This presents no conceptual novelty, but it does involve some technical aspects worth addressing.

In two spatial dimensions the comoving coordinates are described by two scalar fields for each species of solid, $\phi^I_\alpha(x)$ with $I=1,2$ and $\alpha=A,B$, so that at equilibrium $\langle\phi^I_\alpha(x)\rangle=x^I$. 
This again breaks boosts and spatial translations, but also spatial rotations. If, besides homogeneity, one restricts oneself to solids that are isotropic at large distances, then it is necessary to impose an internal $ISO(2)$ symmetry~\cite{Endlich:2012pz}. Under this, the comoving coordinates transform as $\phi_\alpha^I\to O^{IJ}_{\alpha}\phi_\alpha^J+c_\alpha^I$, where $c_\alpha^I$ is a constant vector and $O_\alpha^{IJ}$ a constant $SO(2)$ matrix. This internal group is again spontaneously broken, but a diagonal combination of it with the spacetime Euclidean group is preserved, $ISO(2,1)\times ISO_A(2)\times ISO_B(2)\to \mathbb{R}_t\times ISO(2)$.

The phonon field is again introduced as $\phi_\alpha^I(x)=x^I+\pi_\alpha^I(x)$. If we define $B_{\alpha\beta}^{IJ}\equiv \partial_\mu \phi_\alpha^I \partial^\mu \phi_\beta^J$ and $\Sigma^{IJ}\equiv(\phi_A^I-\phi_B^I)(\phi_A^J-\phi_B^J)$, there are eight operators, $X_i$, built out of them and invariant under the full group, and three operators, $\Delta_i$, invariant only under the diagonal subgroup,
i.e., characterizing the explicit breaking of the symmetry. 
The explicit expressions of the invariants will not be important in the following. More details are reported in Appendix~\ref{app:2Dinv}.
The phonon action, $S=\int d^3x \, F(X_i,\Delta_i)$,  can again be determined by the nonlinear response of the system to shear and stress, as discussed in the previous section. 

We can now expand the action up to cubic order in small fluctuations. Moreover, we perform a field redefinition, $\bm{\pi}_\alpha=O_{\alpha\beta}\mathcal{S}_{\beta\gamma}\bm{\chi}_\gamma$, where $\mathcal{S}$ is a matrix that brings the kinetic term to its canonical form, and $O$ an orthogonal matrix that diagonalizes the mass term. The result is
\begin{widetext}
\begin{align} \label{eq:S2D}
\begin{split}
S&\supset\int d^3x\bigg[\frac{1}{2}\dot{\bm{\chi}}_\alpha^2-\frac{1}{2}K^{(1)}_{(\alpha\beta)}\bm{\nabla}\cdot\bm{\chi}_\alpha\bm{\nabla}\cdot\bm{\chi}_\beta -\frac{1}{2}K_{(\alpha\beta)}^{(2)}\nabla^i\chi_\alpha^j\nabla^i\chi_\beta^j-\frac{1}{2}M^{2}_{\;\;(\alpha\beta)}\bm{\chi}_\alpha\cdot\bm{\chi}_\beta +\lambda^{(1)}_{(\alpha\beta)\gamma}\dot{\bm{\chi}}_\alpha\cdot\dot{\bm{\chi}}_\beta\bm{\nabla}\cdot\bm{\chi}_\gamma \\
& \quad +\lambda_{\alpha\beta\gamma}^{(2)}\dot{\chi}^i_\alpha\dot{\chi}^j_\beta\nabla^i\chi^j_\gamma +\lambda^{(3)}_{(\alpha\beta\gamma)}\bm{\nabla}\cdot\bm{\chi}_\alpha \bm{\nabla}\cdot\bm{\chi}_\beta \bm{\nabla}\cdot\bm{\chi}_\gamma + \lambda^{(4)}_{(\alpha\beta)\gamma}\nabla^i\chi^j_\alpha \nabla^i\chi_\beta^j \bm{\nabla}\cdot\bm{\chi}_\gamma +\lambda^{(5)}_{\alpha\beta\gamma}\nabla^i\chi^j_\alpha \nabla^j\chi^i_\beta \bm{\nabla}\cdot\bm{\chi}_\gamma \\
& \quad + \lambda^{(6)}_{\alpha\beta\gamma}\nabla^i\chi^j_\alpha \nabla^i\chi^k_\beta \nabla^j\chi^k_\gamma + \lambda^{(7)}_{\alpha\beta\gamma}\nabla^i\chi^j_\alpha \nabla^j\chi^k_\beta \nabla^k\chi^i_\gamma + \lambda^{(8)}_{(\alpha\beta)\gamma}\bm{\chi}_\alpha\cdot\bm{\chi}_\beta\bm{\nabla}\cdot\bm{\chi}_\gamma + \lambda^{(9)}_{\alpha\beta\gamma}\chi^i_\alpha\chi^j_\beta\nabla^i\chi^j_\gamma\bigg] \,,
\end{split}
\end{align}
\end{widetext}
with $M^2=\text{diag}(0,\delta^2)$. As in the 1D case, the gap $\delta$ and the couplings $\lambda^{(8)}$ and $\lambda^{(9)}$ vanish with vanishing $\Delta_i$ dependence.
The expressions for the effective coefficients in terms of the derivatives of the energy density are cumbersome but are obtained straightforwardly, as in the 1D case, with simple linear algebra starting from the original Lagrangian.

To read off the spectrum, it is convenient to split the fields into longitudinal and transverse components, $\bm{\chi}_\alpha=\bm{\ell}_\alpha+\bm{t}_\alpha$, such that $\nabla^i\ell_\alpha^j=\nabla^j\ell_\alpha^i$ and $\bm{\nabla}\cdot\bm{t}_\alpha =0$. Looking at the quadratic action, it is simple to show that the dispersion relations are
\begin{subequations}
\begin{align}
\omega^2_{{\sss \text{L}},\text{ac}}=c_{{\sss \text{L}}}^2q^2\,, \qquad\; &\omega^2_{{\sss \text{T}},\text{ac}}=c_{{\sss \text{T}}}^2q^2\,, \label{eq:wac} \\
\omega^2_{{\sss \text{L}},\text{ps}}=\delta^2+\gamma_{{\sss \text{L}}}q^2\,, \qquad\;& \omega^2_{{\sss \text{T}},\text{ps}}=\delta^2+\gamma_{{\sss \text{T}}}q^2\,, \label{eq:wps}
\end{align}
\end{subequations}
with $c_{{\sss \text{L}}}^2\equiv K_{\sss AA}^{(1)}+K_{\sss AA}^{(2)}$, $c_{{\sss \text{T}}}^2\equiv K_{\sss AA}^{(2)}$, $\gamma_{{\sss \text{L}}}\equiv K_{\sss BB}^{(1)}+K_{\sss BB}^{(2)}$ and $\gamma_{{\sss \text{T}}}\equiv K_{\sss BB}^{(2)}$.

Interestingly, the longitudinal and transverse pseudo-acoustic modes share the same gap. This is a consequence of the isotropic approximation, which forces the mass term to be proportional to the identity, $(M^2)_{\alpha\beta}^{ij}=M^2_{\;\;\alpha\beta}\delta^{ij}$. Nonetheless, this is also what happens, for example, in bilayer graphene~\cite{PhysRevB.88.035428}, which features a hexagonal symmetry. Indeed, in two spatial dimensions, when the discrete rotation symmetry is larger than cubic, the quadratic action is identical to that of a fully isotropic material.\footnote{In two dimensions, the only two- and four-index tensors invariant under  discrete rotations larger than cubic are precisely the tensor structures appearing in the quadratic part of Eq.~\eqref{eq:S2D}~\cite{Kang:2015uha}. }


\section{Phonon Decay in 2D}

We now employ our EFT to analytically compute the decay rates of both acoustic and pseudo-acoustic phonons in the large wavelength limit, which are directly related to the thermal conductivity of the material~\cite{srivastava2019physics}. In particular, we can quantize the $\bm{\chi}_\alpha$ fields and use them to evaluate the relativistic matrix elements starting from action~\eqref{eq:S2D}. The phase space is the standard relativistic one~\cite{srednicki2007quantum}. The canonical quantization of the theory, as well as the corresponding Feynman rules, are reported in Appendix~\ref{app:quantization}.

Let us start with the decay of an acoustic phonon, which, for a 2D material, requires some care.
If the dispersion relations \eqref{eq:wac} were exactly linear, the decay of a longitudinal (transverse) acoustic phonon into two longitudinal (transverse) acoustic phonons would produce exactly collinear decay products. In two spatial dimensions, the phase space for this configuration is singular since it diverges as $\sim 1/\theta$, with $\theta$ the angle between the decay products. Now, while the matrix element for the ${{\footnotesize\text{T}},\text{ac}}\to{{\footnotesize \text{T}},\text{ac}}+{{\footnotesize\text{T}},\text{ac}}$ decay vanishes for $\theta=0$, the one for the ${{\footnotesize \text{L}},\text{ac}}\to{{\footnotesize \text{L}},\text{ac}}+{{\footnotesize \text{L}},\text{ac}}$ decay does not, making the total rate formally divergent. This is cured by recalling that the dispersion relation is not exactly linear, but it features a nontrivial curvature: $\omega_{{\sss \text{L}},\text{ac}}(q)=c_{{\sss \text{L}}}q\big(1+\epsilon q^2 + \dots\big)$,
with $|\epsilon| q^2\ll 1$ at small momenta. From the EFT viewpoint, such a (small) curvature is due to operators with a higher number of derivatives, which we did not include in the lowest order action~\eqref{eq:S2D}. {Note that the sign of $\epsilon$ is not fixed a priori, as it depends on the microscopic physics of the system under consideration. For $\epsilon>0$, }
energy and momentum conservation forces the angle between the two outgoing phonons to be $\theta\simeq\sqrt{6\epsilon}(q-q_1)$, where $\bm{q}$ is the momentum of the decaying phonon and $\bm{q}_1$ that of one of the final products. Therefore, at small momenta the ${{\footnotesize \text{L}},\text{ac}}+{{\footnotesize \text{L}},\text{ac}}$ channel dominates the decay rate. {For $\epsilon<0$, instead, the channel with identical initial and final state particles are kinematically forbidden, and the rate does not present any $1/\theta$ divergence. This last case does not present any novelty with respect to what  has been studied previously~\cite{Nicolis:2015sra}, and we will not discuss it here.}

{For positive $\epsilon$, } one can use action~\eqref{eq:S2D} to compute the spin-averaged total decay rate\footnote{Given the derivative couplings, at small momenta the two-body decay dominates the total width.} for an acoustic phonon of initial momentum $\bm q$:
\begin{widetext}
\begin{align} \label{eq:Gammaac}
\begin{split}
\Gamma_\text{ac} = \sqrt{\frac{3}{2\epsilon}} \frac{\bigg[ \sum_{i=3}^{7}\lambda_{\sss AAA}^{(i)} + \big(\lambda_{\sss AAA}^{(1)} + \lambda_{\sss AAA}^{(2)} \big)c_{{\sss \text{L}}}^2 + \big(\lambda_{\sss AAB}^{(8)}-\lambda_{\sss ABA}^{(8)}+\lambda_{\sss AAB}^{(9)}-\frac{1}{2}\lambda_{\sss ABA}^{(9)}-\frac{1}{2}\lambda_{\sss BAA}^{(9)}\big)\frac{K_{\sss AB}^{(1)}+K_{\sss AB}^{(2)}}{\delta^2}\bigg]^2}{32\pi c_{\sss \text{L}}^4} \, q^3\,. 
\end{split}
\end{align}
\end{widetext}

Two comments about the previous expression are in order. 
First, note that when the gap grows, the third term in parenthesis can be neglected, and the rate becomes what one would obtain from the EFT for a single acoustic phonon~\cite{Endlich:2012pz}, which is in agreement with the idea that the pseudo-acoustic one can be integrated out at large gap. Second, because of the considerations made above, {when $\epsilon>0$} the decay width of acoustic phonons in two spatial dimensions is less suppressed at small momenta than what one would expect from naive scaling which, instead, would suggest a $q^4$ behavior. This is a consequence of the well-known extra infrared divergences arising in low-dimensional systems. 

Note also that our analysis applies to an ideal two-dimensional system with no embedding space since it involves only in-plane phonons. Out-of-plane modes~\cite{le2018anomalous} have been shown to have peculiar properties in the absence of external strain, and to contribute sensibly to the decay rate of an acoustic phonon, modifying the behavior at small momenta from $q^3$ to $q^0$~\cite{mariani2008flexural,bonini2012acoustic}. Their dispersion relation in the absence of strain is, in fact, gapless but quadratic, and they consequently contribute to a large part of the available phase space. Nonetheless, it has also been shown that, applying an arbitrarily small strain, the dispersion relation quickly approaches linearity, and a scaling of the decay rate like the one in Eq.~\eqref{eq:Gammaac} is observed~\cite{bonini2012acoustic}.

We can now compute  the spin-averaged decay rate for the pseudo-acoustic phonon. 
To simplify the expression we consider a 2D solid in which the two species, $A$ and $B$, are identical, physically relevant to a description of bilayer graphene. In this case the Lagrangian is of the kind~\eqref{eq:Fweak}, i.e., $F_\text{weak} = f_A + f_B + \delta f$ with $f_A=f_B$, and, consequently, all couplings are symmetric under $A\leftrightarrow B$ and $K_{AB}^{(i)}=0$.
The decay rate for a pseudo-acoustic phonon at rest is found to be
\begin{align}
    \Gamma_\text{ps} = \frac{ \left( 2 g_{\sss ABA}^{(8)} - g_{\sss ABA}^{(9)} \right)^2 }{16 c_{\sss \text{L}} c_{\sss \text{T}} (c_{\sss \text{L}}+c_{\sss \text{T}})^2 }  + O\big(\delta f^3\big)\,.
\end{align}
Note the interesting fact that, to this order in small explicit breaking, the decay rate is independent on the gap itself.\footnote{Interestingly, this is different from what happens in holographic realizations of explicit breaking of translations~\cite{Amoretti:2018tzw}, where the decay width of the quasiparticle is proportional to the gap itself.}


\section{Conclusions}

In this work we presented a relativistic effective field theory for the description of the low-energy degrees of freedom of a solid made of two species, weakly coupled to each other. In this regime the system features two distinct types of excitations: acoustic and pseudo-acoustic phonons. The first are the Goldstone bosons associated with the spontaneous breaking of spacetime symmetries and are consequently gapless. The second are instead pseudo-Goldstone bosons and are characterized by a small gap arising from a perturbative explicit breaking operator.

An EFT formulation of the problem has the important advantage of putting on firm ground several properties of these collective excitations by connecting them to those universal features of the system that depend only on low-energy/large distance physics. It also allows analytical control over the observables, which can be computed for a large class of solids, only via symmetry arguments, and also taking advantage of high energy physics tools.

There are several open questions of both conceptual and phenomenological relevance. One of them is to understand the nature of out-of-plane modes from a low-energy perspective. These modes, as already commented,
 contribute  an important fraction of the total decay rate of acoustic phonons~\cite{bonini2012acoustic} and, in turn, to the thermal conductivity of two-dimensional materials~\cite{srivastava2019physics}. For a first analysis in this direction see~\cite{Coquand:2019yjf}. Similarly, it would be interesting to understand what the contribution of pseudo-acoustic phonons to the thermal conductivity is. 
 Indeed, while optical phonons are typically neglected because of their large gap, pseudo-acoustic phonons have a perturbative gap and could thus contribute  a sizable fraction.
 It would also be interesting to understand if our action~\eqref{eq:S2D} captures any of the features of true optical phonons, despite their large gap. Finally, 
 {we mention one further potential  application of our theory to bilayer graphene. The present construction can be generalized to preserve only a discrete subgroup of rotations, and to include fermionic degrees of freedom for a description of the electron-phonon coupling. These generalizations could provide an EFT that gives a description of (and potentially reveal new insights on) the origin of magic angle superconductivity in bilayer graphene (see, e.g.,~\cite{bistritzer2011moire,cao2018unconventional,cao2018correlated,lu2019superconductors}).}
 We leave these questions for future work.


\begin{acknowledgments}
We are grateful to M.~Baggioli, G.~Cuomo, A.~Khmelnitsky, A.~Nicolis, R.~Penco, and R.~Rattazzi for the important discussions. We are especially thankful to G.~Chiriac\`o for several enlightening conversations and comments on the draft.
A.E. is supported by the Swiss National Science Foundation under Contract No. 200020-169696 and through the National Center of Competence in Research SwissMAP. E.G. acknowledges support from the International Max Planck Research School for Precision Tests of Fundamental Symmetries in Particle Physics, Nuclear Physics, Atomic Physics and Astroparticle Physics.
T.M. is supported by the World Premier International Research Center Initiative (WPI) MEXT, Japan, and by JSPS KAKENHI Grants No. JP18K13533, No. JP19H05810, No. JP20H01896, and No. JP20H00153.
\end{acknowledgments}


\appendix

\section{Effective couplings for the 1D solid} \label{app:couplings}

Here we report the explicit expressions for the couplings and parameters of the EFT for the 1D case, written in terms of the derivatives of the Lagrangian evaluated on the background configuration, $\langle\phi\rangle=x$. The subscripts indicate derivatives with respect to a given invariant. The effective couplings appearing in action~\eqref{eq:S21D} are
\begin{align}
    g_{\sss AA}&=F_{\sss X_1}\,, \quad g_{\sss BB}=F_{\sss X_2}\,, \quad g_{\sss AB}=\frac{1}{2}F_{\sss X_3}\,, \notag \\
    g_{\sss AA}^\prime&=F_{\sss X_1}+2F_{\sss X_1X_1}+\frac{1}{2}F_{\sss X_3X_3}+2F_{\sss X_1X_3}\,, \notag \\
    g_{\sss BB}^\prime&=F_{\sss X_2}+2F_{\sss X_2X_2}+\frac{1}{2}F_{\sss X_3X_3}+2F_{\sss X_2X_3}\,, \\
    g_{\sss AB}^\prime&=\frac{1}{2}F_{\sss X_3}+\frac{1}{2}F_{\sss X_3X_3}+2F_{\sss X_1X_2}+F_{\sss X_1X_3}+F_{\sss X_2X_3}\,, \notag \\  
    g_{\sss \Delta}&=F_{\sss \Delta}\, \notag
\end{align}
for the quadratic ones, and
\begin{align}
\begin{split}
    y_{\sss AAA}&=2F_{\sss X_1X_1}+F_{\sss X_1X_3}+\frac{4}{3}F_{\sss X_1X_1X_1} +2F_{\sss X_1X_1X_3} \\
    & \quad +F_{\sss X_1X_3X_3}+\frac{1}{6}F_{\sss X_3X_3X_3}\,, \\
    y_{\sss AAB}&=\frac{2}{3}F_{\sss X_1X_2}+F_{\sss X_1X_3}+\frac{1}{3}F_{\sss X_3X_3} +\frac{4}{3}F_{\sss X_1X_1X_2} \\
    & \quad +\frac{2}{3}F_{\sss X_1X_1X_3}+\frac{4}{3}F_{\sss X_1X_2X_3}  +\frac{2}{3}F_{\sss X_1X_3X_3}\\
    &\quad+\frac{1}{3}F_{\sss X_2X_3X_3}+\frac{1}{6}F_{\sss X_3X_3X_3}\,, \\
    y_{\sss ABB}&=y_{\sss AAB} \;\text{ with }\;X_1\leftrightarrow X_2\,, \\
    y_{\sss BBB}&=y_{\sss AAA}\; \text{ with }\;X_1\leftrightarrow X_2\,, \\
    y_{\sss AAA}^\prime&=2F_{\sss X_1X_1}+F_{\sss X_1X_3}\,,  \\
    y_{\sss AAB}^\prime&=2F_{\sss X_1X_2}+F_{\sss X_1X_3}\,, \\
    y_{\sss ABA}^\prime&=F_{\sss X_1X_3}+\frac{1}{2}F_{\sss X_3X_3}\,, \\
    y_{\sss ABB}^\prime&=y_{\sss ABA}^\prime\;\text{ with }\;X_1\leftrightarrow X_2\,, \\
    y_{\sss BBA}^\prime&=y_{\sss AAB}^\prime \; \text{ with }\; X_1\leftrightarrow X_2\,, \\
    y_{\sss BBB}^\prime&=y_{\sss AAA}^\prime \; \text{ with }\; X_1\leftrightarrow X_2\,, \\
    y^{\prime\prime}_{\sss A}&=2F_{\sss X_1\Delta}+F_{\sss X_3\Delta}\,, \\
    y^{\prime\prime}_{\sss B}&=y^{\prime\prime}_{\sss A} \; \text{ with }\; X_1\leftrightarrow X_2
\end{split}
\end{align}
for the cubic ones.

The parameters appearing in the dispersion relation for the acoustic and optical phonons in one spatial dimension, Eq.~\eqref{eq:dispersion1D}, are instead
\begin{align}
    c_s^2&=\frac{g_{\sss AA}^\prime+g_{\sss BB}^\prime+2g_{\sss AB}^\prime}{g_{\sss AA}+g_{\sss BB}+2g_{\sss AB}}\,, \notag \\
    \gamma&=\frac{1}{(g_{\sss AA}+g_{\sss BB}+2g_{\sss AB})(g_{\sss AA}g_{\sss BB}-g_{\sss AB}^2)} \times \notag \\ 
    & \quad \times \bigg[g_{\sss AA}^2g_{\sss BB}^\prime +g_{\sss BB}^2g_{\sss AA}^\prime-2g_{\sss AA}g_{\sss BB}g_{\sss AB}^\prime \\
    &\quad + g_{\sss AB}^2\big(g_{\sss AA}^\prime+g_{\sss BB}^\prime-2g_{\sss AB}^\prime\big) \notag \\
    & \quad +2g_{\sss AA}g_{\sss AB}\big(g_{\sss BB}^\prime-g_{\sss AB}^\prime\big) +2g_{\sss BB}g_{\sss AB}\big(g_{\sss AA}^\prime-g_{\sss AB}^\prime\big)\bigg]\,. \notag
\end{align}


\section{2D invariants} \label{app:2Dinv}

Here we collect all the independent operators that are invariant under the symmetry group relevant for two dimensional solids with two atomic species. After first imposing Poincar\'e and shift invariance one obtains the following matrices:
\begin{align} \label{eq:Bmatrices}
&B_{AA}^{IJ}=\partial_\mu\phi_A^I\partial^\mu\phi_A^J\,,\;\;\; B_{BB}^{IJ}=\partial_\mu\phi_B^I\partial^\mu\phi_B^J\,, \\
&B_{AB}^{IJ}=\partial_\mu\phi_A^I\partial^\mu\phi_B^J\,, \;\;\; \Sigma^{IJ}=(\phi_A^I-\phi_B^I)(\phi_A^J-\phi_B^J)\,. \notag
\end{align}
The $B_{\alpha\beta}$ matrices transform linearly under the initial $SO_A(2)\times SO_B(2)$ group---i.e., $B_{\alpha\beta}\to O_\alpha \cdot B_{\alpha\beta}\cdot O_\beta^T$, where $O_\alpha$ is an orthogonal matrix belonging to $SO_\alpha(2)$. The matrix $\Sigma$, instead, transforms linearly only under the final unbroken $SO(2)$ and is invariant only under the unbroken $U(1)$.

The matrices in Eq.~\eqref{eq:Bmatrices} have a total of 12 components. However, we can always perform an unbroken $SO(2)$ rotation to bring the number down to 11, which is the the number of independent invariants under spatial and internal translations.
The following eight are invariant under the full $SO_A(2)\times SO_B(2)$ group and hence are compatible with its spontaneous breaking:
\begin{align} \label{eq:inv1}
    X_1 &= \left[B_{AA}\right]\,, \;\; X_2 = \left[B_{BB}\right]\,, \;\; X_3 = \left[B_{AB}B_{AB}^T\right]\,, \notag \\
    X_4 &= \left[B_{AA}^2\right]\,, \;\; X_5 = \left[B_{BB}^2\right]\,, \;\;  X_6 = \left[(B_{AB}B_{AB}^T)^2\right]\,, \notag \\
    X_7 &= \left[B_{AB}B_{AB}^T B_{AA}\right]\,, \;\; X_8 = \left[B_{AB}^T B_{AB} B_{BB}\right]\,.
\end{align}
Here we represent the trace with $[\,\cdots]$ . The remaining three operators are invariant only under the unbroken group and hence explicitly break the initial symmetry:
\begin{align} \label{eq:inv2}
\begin{split}
    \Delta_1 = \left[B_{AA}B_{BB}\right]\,, \quad \Delta_2 = \left[ \Sigma \right]\,, \\
    \Delta_3=\frac{\left[\Sigma B_{AA}\right]+\left[\Sigma B_{BB}\right]}{2}-\left[\Sigma\right]\,.
\end{split}
\end{align}
The expression for $\Delta_3$ has been chosen so that it does not contribute to the quadratic action.


\section{Canonical quantization and Feynman rules for the 2D solid} \label{app:quantization}

First, we provide the details of the canonical quantization for the fields $\bm{\chi}_\alpha$ appearing in action~\eqref{eq:S2D}. Following the standard canonical quantization procedure, we expand them in creation and annihilation operators and require that they satisfy the equations of motion. Since the quadratic action is in general nondiagonal, $\bm{\chi}_A$ and $\bm{\chi}_B$ obey a set of coupled linear differential equations, and therefore both of them contain creation/annihilation operators for the acoustic and pseudo-acoustic phonons. We thus write
\begin{align} \label{eq:canonicalfield}
   \bm\chi_\alpha(x)  =  \sum_{\sss \lambda,f}  \int  \frac{d^2q}{(2\pi)^2 2\omega_{\sss \lambda f}}  \bm{\epsilon}^{\sss \lambda}_{\bm q} C^{\sss \lambda f}_{\alpha}(q) a^{\sss \lambda f}_{\bm q} e^{iq_{\sss \lambda f} \cdot  x} + \text{H.c.} \,, \notag
\end{align}
where $\lambda$ is the phonon's polarization (longitudinal/transverse) and $f$ its ``flavor'' (acoustic/pseudo-acoustic), and $a_{\bm q}^{\scriptscriptstyle \lambda f}$ is the annihilation operator, normalized so that
\begin{align}
    \left[ a_{\bm q}^{\sss \lambda f} , \big(a_{\bm q'}^{\sss \lambda' f'}\big)^\dagger  \right] =  2\omega_{\sss \lambda f}(2\pi)^2\delta^{(2)}(\bm q-\bm q')\delta^{\sss \lambda\lambda'}\delta^{\sss ff'}\,.
\end{align}
Moreover, $\bm{\epsilon}_{\bm q}^{\scriptscriptstyle \lambda}$ is a polarization vector, which for longitudinal and trasverse phonons is given respectively by $\epsilon_{\bm q}^{{\sss \text{L}},i}=\hat{q}^i$ and $\epsilon_{\bm q}^{{\sss \text{T}},i}=\epsilon^{ij}\hat{q}^j$, such that they satisfy the completeness relation, $\sum_\lambda {\epsilon}_{\bm q}^{\lambda,i}{\epsilon}_{\bm q}^{\lambda,j}=\delta^{ij}$.
To determine the overlap functions, $C_\alpha^{\sss \lambda f}(q)$, we first require the fields to satisfy the equal-time commutation relations, $[\chi_\alpha^i(t,\bm x),\dot\chi_\alpha^j(t,\bm y)]=i\delta^{(2)}(\bm x-\bm y)\delta^{ij}$, leading to
\begin{align}
    \sum_{\sss f} \left| C_\alpha^{\sss \text{L} f}\right|^2 = \sum_{\sss f} \left| C_\alpha^{\sss \text{T} f}\right|^2 = 1\,.
\end{align}
We then impose the linear equations of motion and obtain, up to order $O(q^2)$,
\begin{align}
\begin{split}
&C_{\sss A}^{{\sss \text{L}},\text{ac}}=C_{\sss A}^{{\sss \text{T}},\text{ac}}=1\,,  \quad C_{\sss B}^{{\sss \text{L}},\text{ac}}=-\frac{K_{\sss AB}^{(1)}+K_{\sss AB}^{(2)}}{\delta^2}q^2\,, \\
&C_{\sss B}^{{\sss \text{T}},\text{ac}}=-\frac{K_{\sss AB}^{(2)}}{\delta^2}q^2\,, \quad C_{\sss A}^{{\sss \text{L}},\text{ps}}= \big|C_{\sss B}^{{\sss \text{L}},\text{ac}}\big| \,, \\
&C_{\sss A}^{{\sss \text{T}},\text{ps}}=\big|C_{\sss B}^{{\sss \text{T}},\text{ac}}\big|\,, \quad C_{\sss B}^{{\sss \text{L}},\text{ps}}=\text{sgn}\left(K_{\sss AB}^{(1)}+K_{\sss AB}^{(2)}\right)\,,  \\
&C_{\sss B}^{{\sss \text{T}},\text{ps}}=\text{sgn}\,K_{\sss AB}^{(2)}\,.
\end{split}
\end{align}

We next present the Feynman rules. From the canonical expression for the $\bm{\chi}_\alpha$ field we can deduce the propagator, $G_{\alpha\beta}^{ij}(x)=\langle \Omega |T\chi_\alpha^i(x)\chi_\beta^j(0)|\Omega \rangle$, where $|\Omega \rangle$ is the interacting solid vacuum, and $T$ enforces the time-ordered product. With standard quantum field theory techniques we deduce the following rules:
\begin{align*}
    \!\includegraphics[width=0.21\columnwidth,valign=c]{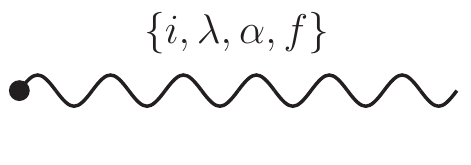} =&\; \epsilon_{\bm{q}}^{\lambda,i}C_\alpha^{\sss \lambda f}(q) \qquad (\text{external leg}) \\
    \!\includegraphics[width=0.27\columnwidth,valign=c]{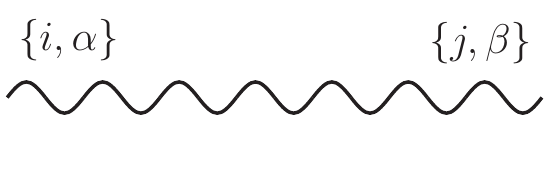} =&\sum_{\sss \lambda, f}C_\alpha^{\sss \lambda f}(q)C_\beta^{\sss \lambda f}(q)\frac{i \epsilon_{\bm{q}}^{\lambda,i} \epsilon_{\bm{q}}^{\lambda,j}}{\omega^2-\omega_{\sss \lambda f}^2(q)+i\varepsilon} \\
    \!\includegraphics[width=0.29\columnwidth,valign=c]{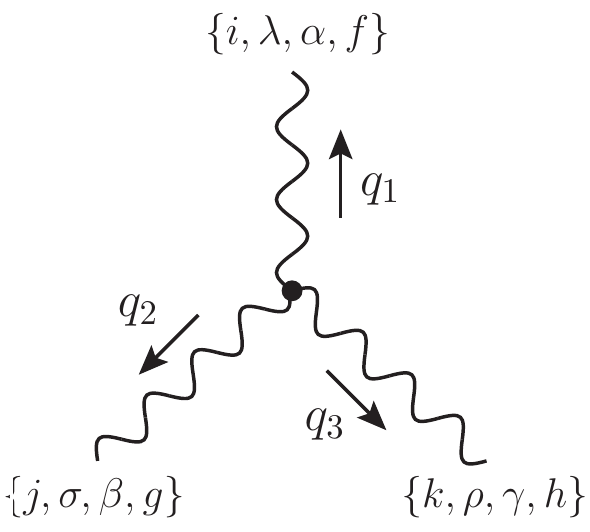} \!\!\!\!\!=& -\lambda_{(\alpha\beta)\gamma}^{(1)}\omega_{\sss \lambda f}(q_1)\omega_{\sss \sigma g}(q_2)\delta^{ij}q_3^k \\[-2.5em]
    &-\lambda_{\alpha\beta\gamma}^{(2)}\omega_{\sss \lambda f}(q_1)\omega_{\sss \sigma g}(q_2)\delta^{jk}q_3^i \\
    &-\lambda_{(\alpha\beta\gamma)}^{(3)}q_1^i q_2^j q_3^k -\lambda_{(\alpha\beta)\gamma}^{(4)}\bm{q}_1\cdot\bm{q}_2\delta^{ij}q_3^k \\
    &-\lambda_{\alpha\beta\gamma}^{(5)}q_1^jq_2^iq_3^k -\lambda_{\alpha\beta\gamma}^{(6)}\bm{q}_1\cdot\bm{q}_2 \delta^{jk} q_3^i \\
    &-\lambda_{\alpha\beta\gamma}^{(7)}q_1^kq_2^iq_3^j + \lambda_{(\alpha\beta)\gamma}^{(8)}\delta^{ij}q_3^k \notag \\
    &+\lambda_{\alpha\beta\gamma}^{(9)}\delta^{jk}q_3^i + \text{permutations}\,,
\end{align*}
where we recall that the indices $i,j,k$ run over spatial components, $\lambda,\sigma,\rho$ run over longitudinal/transverse polarizations, $\alpha,\beta,\gamma$ run over the solid labels $A$ and $B$, and finally $f,g,h$ run over the acoustic/pseudo-acoustic flavors. Moreover, by ``permutations'' we mean the possible combinations of the collective indices $\{q_1,i,\lambda,\alpha,f\}$, $\{q_2,j,\sigma,\beta,g\}$ and $\{q_3,k,\rho,\gamma,h\}$. It is possible to check that the propagator is indeed the inverse of the kinetic matrix of action~\eqref{eq:S2D}.

\bibliographystyle{apsrev4-1}
\bibliography{biblio}

\end{document}